\documentstyle[aps,epsfig,twocolumn]{revtex}
\newcommand{\sn}{\,\mathrm{sn}\,}
\newcommand{\tinyhalf}{{\scriptstyle {1 \over 2}}}

\begin{document}
\draft \small \twocolumn[\hsize\textwidth\columnwidth\hsize\csname
@twocolumnfalse\endcsname \title{Exact solutions of the two-mode
model of multicomponent Bose-Einstein condensates}

\author{Vadym Vekslerchik\cite{email1} and V\'{\i}ctor M. P\'erez-Garc\'{\i}a\cite{email2}}

\address{Departamento de Matem\'aticas, Escuela T\'ecnica Superior de
  Ingenieros Industriales\\ Universidad de Castilla-La Mancha, 13071 Ciudad
  Real, Spain }

\date{\today}

\maketitle


\begin{abstract}
  We find the explicit solution of a two--mode model used to
  explain vortex dynamics in multicomponent Bose-Einstein
  condensates. We prove that all the solutions are constants or
  periodic functions and give explicit formulae for
  the time evolution of the populations of the two atomic species present in the
  condensate.
\end{abstract}

 ]


\narrowtext


\section{Introduction}

One of the most remarkable achievements of contemporary physics
has been the realization of Bose-Einstein condensation with
ultracold atomic gases \cite{experim}. In those experiments
bosonic neutral gases where cooled down below the critical
temperature and a collective coherent behavior was observed in the
gas cloud.

One later interesting results in this field were the achievement
of simultaneous condensation of several atomic species using
sympathetic cooling \cite{Sympathetic} and the generation of
multiple condensates using r.f. transitions between different
hyperfine levels of Rb$^{87}$ \cite{2comp}. These two species are
usually denoted by $|1\rangle$ and $|2\rangle$. Despite its
complexity, this system may be described in the mean field limit
using two macroscopic wavefunctions describing each of the atomic
species. After these achievements  many theoretical and
experimental works on multicomponent systems followed
\cite{gasou}.

The first generation of vortices in Bose-Einstein condensates
reported in Ref. \cite{MMatt99}, has been one of the most
remarkable goals which has been achieved using multicomponent
systems.  From Ref.\cite{MMatt99} we know that while one may build
two possible configurations with a unit charge vortex only one of
them is stable.  The stable configuration corresponds to the
vortex placed on the $|1\rangle$ state, namely the one with the
largest scattering length. The configuration with the vortex
placed in the $|2\rangle$ state, on the other hand, leads to some
kind of instability.

In recent works \cite{PRL3,Skryabin} numerical simulations have
been used to show that the dynamics of the unstable configuration
can be understood within the framework of mean field theories for
the double--condensate system. In a later work \cite{PRA3}, it was
shown that a simple two--mode model is enough  to describe
accurately the dynamics of the  multiple--condensate system in two
dimensional setups (e.g. oblate condensates where the
$z$-coordinate may be neglected). The model was able to describe
the most relevant features found in experiment and also in the
mean field theories of the system \cite{PRL3,Skryabin}.

In this paper we concentrate on the analysis of the simple model
reported in Ref. \cite{PRA3} proving that the system is integrable
and all the solutions are periodic or constants (equilibria). This
proves that the tendency of the system to exhibit periodic
exchange of vortices between both atomic species is fundamental
and not the consequence of a fortunate choice of parameters. We
also study the period of the vortex trasfer mechanism as a
function of the physical parameters of the problem, in particular
we find a linear dependence on the total number of particles $N$
of the system.

\section{Two mode model for the multiple species condensate.}

\subsection{Gross-Pitaevskii equations for the two-species
system.}

 In this work we will use the zero temperature
approximation, in which collisions between the condensed and non
condensed atomic clouds are neglected. In the two species case
this leads to a pair of coupled Gross-Pitaevskii equations (GPE)
for the condensate wavefunctions of each species
\begin{mathletters}
\label{GPE}
\begin{eqnarray}
i \hbar \frac{\partial}{\partial t} \Psi_1 &=& \left[ - \frac{\hbar^2\nabla^2}{2m} +
 V_1 + U_{11} |\Psi_1|^2 + U_{12} |\Psi_2|^2
 \right] \Psi_1, \\
i \hbar \frac{\partial}{\partial t} \Psi_2 &=& \left[ - \frac{\hbar^2\nabla^2}{2m} +
  V_2 + U_{21} |\Psi_1|^2 + U_{22} |\Psi_2|^2
 \right] \Psi_2,
\end{eqnarray}
\end{mathletters}
where $U_{ij} = 4\pi \hbar^2 a_{ij}/m$ are constants controlling
the nonlinear behavior, which are proportional to the the $s$-wave
scattering lengths of 1-1, $(a_{11})$, 2-2 $(a_{22})$, and 1-2
$(a_{12})$ binary collisions.

To simplify the formalism we assume that both trapping potentials
are concentric and spherically symmetric, $V_1(\vec{r}) =
V_2(\vec{r}) = \frac{1}{2}m \omega^2 r^2$, just like in the
experiment. The consideration of the differences between $V_1$ and
$V_2$ does not add new physics to the model as discussed in Ref.
\cite{PRA3}.

Next we change to a new set of units based on the trap characteristic length,
$a_0=\sqrt{\hbar/m\omega}$, and period, $\tau=1/\omega$ defined as $x
\rightarrow x/a_0, t \rightarrow t/\tau$, $u_{ij} = 4\pi N_j a_{ij}/a_0$ and
$\Psi_j({\bf x}) = N_j \psi_j({\bf x})$.  Equations (\ref{GPE}) conserve the
number of particles on each hyperfine level and so we may choose
\begin{equation}
\int|\psi_1(\vec{r})|^2= \int|\psi_2(\vec{r})|^2 \equiv  1.
\end{equation}
This choice implies that the particle number of each species
appears in the nonlinear coefficients $u_{ij}$.

With the previous rescaling, the GPE for the multicomponent system read
\begin{mathletters}
\label{simple}
\begin{eqnarray}
i  \frac{\partial}{\partial t} \psi_1 &=& \left[ - \frac{1}{2} +
 \frac{1}{2}r^2 + u_{11} |\psi_1|^2 + u_{12} |\psi_2|^2
 \right] \psi_1, \\
i  \frac{\partial}{\partial t} \psi_2 &=& \left[ - \frac{1}{2} +
  \frac{1}{2}r^2 + u_{21} |\psi_1|^2 + u_{22} |\psi_2|^2
 \right] \psi_2.
\end{eqnarray}
\end{mathletters}

The coefficients of the matrix of nonlinear coefficients satisfy
the relations $u_{11}/u_{12} = a_{11}N_1/a_{12}N_2, u_{21}/u_{22}
= a_{12}N_1/a_{22}N_2$, which means that except for the particular
case in which $N_1 = N_2 = N,$ this matrix is nonsymmetric.  In
terms of the population imbalance $\beta = N_2/N_1,$ and for a
fixed total number of particles the values are $ u_{11} = g
a_{11}/(1+\beta)$, $ u_{12} = g a_{12}\beta/(1+\beta)$, $u_{21} =
g a_{12}/(1+\beta) $, $u_{22} = g a_{22}\beta /(1+\beta)$, where
$g =4\pi N/a_0$.

\subsection{Derivation of a two-mode model}

For completeness we will derive here the two--mode model following
a different formalism than that of Ref. \cite{PRA3}. The idea is
to use a limited expansion

\begin{equation}
\label{app}
  \psi_{n}(t,\vec{r}) =
  \sum_{k=1}^{2} a_{nk}(t) \phi_{k}(\vec{r}),
  \end{equation}
  where we take $\phi_k(\vec{r})$ to be the lowest energy
  solutions of the linear problem
  \begin{equation}
  \label{basis}
  \left[ - {1 \over 2} \Delta + V(\vec{r}) \right] \phi_{k}(\vec{r}) =
  E_{k}\phi_{k}(\vec{r}).
\end{equation}
Although we have choosen a harmonic oscillator basis, any suitable
truncated basis including a mode resembling the ground state and
another one corresponding to a state with a vortex could be used
in this expansion. Despite its simplicity and crudeness (only two
modes of the full expansion are kept), this approximation has been
shown to retain many qualitative features of the dynamics of the
system \cite{PRA3}. Inserting Eq. (\ref{app}) with $\phi$ given by
(\ref{basis}) into Eq. (\ref{simple}) we get
\begin{equation}
  \sum_{k=1}^{2}
    \left[-i \dot a_{nk}(t) + E_{k}a_{nk}(t) + U_{n}(t,\vec{r})a_{nk}(t)
    \right]
    \phi_{k}(\vec{r}) = 0,
\end{equation}
where $U_{n}(t,\vec{r}) =  \sum_{m=1}^{2} u_{nm}
\left|\psi_{m}(t,\vec{r}) \right|^{2}$. After some algebra we find
\begin{equation}
  \partial_{t} \left|a_{nk} \right|^{2} =
  - 2 u_{n\bar{n}} C_{12} \left| \, a_{11} a_{12} a_{21} a_{22} \, \right|
  \sin\Phi_{nk},
\end{equation}
where $  C_{jk} = \left<  \left|\phi_{j} \right|^{2}
\left|\phi_{k} \right|^{2} \right>$ and $  \Phi \equiv \Phi_{11} =
- \Phi_{12} = - \Phi_{21} = \Phi_{22} =
  \arg ( a_{11} a_{22} / a_{12} a_{21} )$.
The equations for the phases are
\begin{equation}
  \partial_{t} \arg a_{nk} =
  - E_{k}
  - {\mathrm{Re}} \;
    \sum_{k'=1}^{2} \frac{a_{nk'}}{a_{nk}}
    \left<  \overline{\phi_{k}} \; U_{n} \; \phi_{k'} \right>.
\end{equation}
where $\left< A \right> = \int d^2\vec{r} A$. Calculating $ \left<
\overline{\phi_{k}} \; U_{n} \; \phi_{k'} \right>$ for the two
cases $k' = k$ and $k' = \overline{k}$, where $\overline{k}$ is
the complementary value of the index, ${k,k'} = {1,2}$
\begin{eqnarray}
\left< \overline{\phi_{k}} \; U_{n} \;   \phi_{k} \; \right> & = &
\sum_{m=1}^{2}
    u_{nm}
    \left\{
      C_{kk} \left| a_{mk} \right|^{2} +
      C_{12} \left| a_{m\bar{k}} \right|^{2}
    \right\}, \\
      \left<  \overline{\phi_{k}}\; U_{n} \; \phi_{\bar{k}} \; \right>  &
      = &
  C_{12}
  \sum_{m=1}^{2} u_{nm} \; a_{mk} \; \overline{a_{m\bar{k}}},
  \end{eqnarray}
we get the following set of equations for our problem
\begin{eqnarray}
 \partial_{t} \Phi & = & - \Gamma -
  C_{12} \; \rho_{11} \rho_{12} \rho_{21} \rho_{22} \cos \Phi \times
  \nonumber\\
 & &  \left[
    u_{12} \left( { 1 \over \rho^{2}_{11} } - { 1 \over \rho^{2}_{12} } \right) +
    u_{21} \left( { 1 \over \rho^{2}_{22} } - { 1 \over \rho^{2}_{21} } \right)
  \right],
\label{dphi} \\
  \partial_{t} \rho^{2}_{11} &=&
  - 2 C_{12} u_{12} \, \rho_{11} \rho_{12} \rho_{21} \rho_{22} \,
  \sin\Phi,
\label{dr11}
\\
  \partial_{t} \rho^{2}_{12} &=&
  + 2 C_{12} u_{12} \, \rho_{11} \rho_{12} \rho_{21} \rho_{22} \,
  \sin\Phi,
\label{dr12}
\\
  \partial_{t} \rho^{2}_{21} &=&
  + 2 C_{12} u_{21} \, \rho_{11} \rho_{12} \rho_{21} \rho_{22} \,
  \sin\Phi,
\label{dr21}
\\
  \partial_{t} \rho^{2}_{22} &=&
  - 2 C_{12} u_{21} \, \rho_{11} \rho_{12} \rho_{21} \rho_{22} \,
  \sin\Phi,
\label{dr22}
\end{eqnarray}
where
\begin{equation}  a_{nk} = \rho_{nk}
\exp\left(i\theta_{nk}\right),
\end{equation}
and
\begin{equation}
\label{defgamma} \Gamma =  \sum_{nk} \gamma_{nk} \rho^{2}_{nk},
\end{equation}
being
\begin{eqnarray}
  \gamma_{11} &=&
    C_{11} \left( u_{11} - u_{21} \right) +
    C_{12} \left( u_{21} - 2u_{11} \right),
\\
  \gamma_{12} &=&
    C_{22} \left( u_{21} - u_{11} \right) +
    C_{12} \left( 2u_{11} - u_{21} \right),
\\
  \gamma_{22} &=&
    C_{22} \left( u_{22} - u_{12} \right) +
    C_{12} \left( u_{12} - 2u_{22} \right),
\\
  \gamma_{21} &=&
    C_{11} \left( u_{12} - u_{22} \right) +
    C_{12} \left( 2u_{22} - u_{12} \right).
\end{eqnarray}
These equations must be complemented with the appropriate initial
conditions for the densities $\rho_{kl}$ and phase $\Phi$.
\begin{equation}
  \rho_{nk}(t_{0}) = \hat\rho_{nk}
  \qquad
  n,k = 1,2
  \qquad
  \qquad
  \Phi(t_{0}) = \hat\Phi.
\end{equation}
The form of the two-mode model [Eqs. (\ref{dphi})-(\ref{dr22})]
presented here is slightly different from that of Ref.
\cite{PRA3}, but is more suitable for our purposes in this paper.

\section{Solution of the equations.}

Let us first note that the two-mode model equations have the
following constants of motion:
\begin{eqnarray}
  A_{1} &=&
    \rho^{2}_{11} + \rho^{2}_{12} =
    \mathrm{const} =
    {\hat\rho}^{2}_{12} + {\hat\rho}^{2}_{12},
\\
  A_{2} &=&
    \rho^{2}_{21} + \rho^{2}_{22} =
    \mathrm{const} =
    {\hat\rho}^{2}_{21} + {\hat\rho}^{2}_{22},
\\
  L &=&
    u_{21}\rho^{2}_{12} + u_{12}\rho^{2}_{22} =
    u_{21}{\hat\rho}^{2}_{12} + u_{12}{\hat\rho}^{2}_{22}.
\end{eqnarray}
It follows from (\ref{dr11})-(\ref{dr22}) and (\ref{defgamma})
that
\begin{equation}
  \partial_{t} \Gamma =
  2 \gamma_{*} C_{02} \, \rho_{11} \rho_{12} \rho_{24} \rho_{22} \,
  \sin\Phi,
\label{dgamma2}
\end{equation}
where
\begin{equation}
  \gamma_{*} =
  u_{12} \left( \gamma_{12} - \gamma_{11} \right) +
  u_{21} \left( \gamma_{11} - \gamma_{22} \right).
\end{equation}
The following relations are then evident
\begin{eqnarray}
  \rho_{11}^{2} &=&
     \hat\rho_{11}^{2} - { u_{12} \over \gamma_{*} } \left( \Gamma - \hat\Gamma \right)
\label{rsol11}
\\
  \rho_{12}^{2} &=&
      \hat\rho_{12}^{2} + { u_{12} \over \gamma_{*} } \left( \Gamma - \hat\Gamma
      \right),
\label{rsol12}
\\
  \rho_{21}^{2} &=&     \hat\rho_{21}^{2}
  + { u_{21} \over \gamma_{*} } \left( \Gamma - \hat\Gamma
  \right),
\label{rsol21}
\\
  \rho_{22}^{2} &=&
  \hat\rho_{22}^{2} - { u_{21} \over \gamma_{*} } \left( \Gamma - \hat\Gamma
  \right),
\label{rsol22}
\end{eqnarray}
with $\hat\Gamma =  \sum_{nk} \gamma_{nk} \hat\rho^{2}_{nk}$.
These equations imply that all the relevant densities may be
obtained from a single quantity, $\Gamma(t)$. Our goal now will be
to find an equation ruling its dynamics. Let us notice that Eq.
(\ref{dphi}) can be rewritten as
\begin{equation}
  \partial_{t} \, \rho_{11} \rho_{12} \rho_{21} \rho_{22} \cos\Phi =
  { 1 \over 2 C_{12} \gamma_{*} }
  \Gamma \, \partial_{t} \, \Gamma
\end{equation}
and solved:
\begin{equation}
  \rho_{11} \rho_{12} \rho_{21} \rho_{22} \cos\Phi -
  \hat\rho_{11} \hat\rho_{12} \hat\rho_{21} \hat\rho_{22} \cos\hat\Phi
  =
  { 1 \over 4 C_{12} \gamma_{*} } \left( \Gamma^{2} - \hat\Gamma^{2} \right)
\end{equation}
Using the last equation and (\ref{rsol11})-(\ref{rsol22}), Eq.
(\ref{dgamma2}) for $\Gamma_{t}$ can be presented as
\begin{equation}
\label{gammap}
  \Gamma_{t}^{2} = {\mathcal{P}}_{4}(\Gamma)
\end{equation}
where
\begin{eqnarray}
  {\mathcal{P}}_{4}(\Gamma) = - { 1 \over 4 }
    \left[
      \Gamma^{2} - \hat\Gamma^{2} +
      4 \gamma_{*} C_{12} \,
      \hat\rho_{11}\hat\rho_{12}\hat\rho_{21}\hat\rho_{22}
      \cos\hat\Phi
    \right]^{2} + \nonumber \\
  4 C_{12}^{2} \, {  u_{12}^{2} u_{21}^{2} \over \gamma_{*}^{2} } \,
    \left( \Gamma - \Gamma_{11} \right)
    \left( \Gamma - \Gamma_{12} \right)
    \left( \Gamma - \Gamma_{21} \right)
    \left( \Gamma - \Gamma_{22} \right)
    \label{poly}
\end{eqnarray}
and the constants $\Gamma_{nk}$, $n,k=1,2$ are defined as $
\Gamma_{11} = \hat\Gamma + \gamma_{*}/u_{12} \hat\rho_{11}^{2} $,
$
  \Gamma_{12} =
    \hat\Gamma - \gamma_{*}/u_{12} \, \hat\rho_{12}^{2}$,
$  \Gamma_{21} =
    \hat\Gamma - \gamma_{*}/u_{21} \, \hat\rho_{21}^{2}$,
$\Gamma_{22} =
    \hat\Gamma + \gamma_{*}/u_{21} \, \hat\rho_{22}^{2}.$
Thus, the solution of our equations can be obtained from a single
equation for $\Gamma$, which, using the previous equations, allows
to obtain the expressions for the densities $\rho_{jk}(t)$. In
particular Eq. (\ref{gammap}) leads to
\begin{equation}
  t - t_{0} =
  \int\limits_{\hat\Gamma}^{\Gamma(t)}
  { d\Gamma' \over \sqrt{{\mathcal{P}}_{4}(\Gamma')}}.
\end{equation}
Let us rewrite the polinomial (\ref{poly}) in terms of its roots
as follows
\begin{equation}
  {\mathcal{P}}_{4}(\Gamma) =
  {\mathcal{P}}_{*}
    \left( \Gamma - P_{1} \right)
    \left( \Gamma - P_{2} \right)
    \left( \Gamma - Q_{1} \right)
    \left( \Gamma - Q_{2} \right)
\end{equation}
where $  {\mathcal{P}}_{*}=
  4 C_{12}^{2} \, u_{12}^{2} u_{21}^{2}/\gamma_{*}^{2}
  - 1/4 $ and ${P_1,P_2,Q_1,Q_2}$ are the roots of
  ${\mathcal{P}}_4$. There are at least
  two real roots $P_1,P_2$ of ${\mathcal{P}}_4$ satisfying that
  $\hat{\Gamma}$ is always located
 between them, i.e.  $P_{1} < \hat\Gamma < P_{2}$.

 To prove the last affirmation let us first compute
 $$ {\mathcal{P}}_4(\tilde{\Gamma}) = 4 C_{12}^2
 \gamma_*^2\sin^2\Phi > 0.$$
 Next we evaluate
 \begin{eqnarray*}
{\mathcal{P}}_4(\Gamma_{11}) & = & - \frac{1}{4}\left[
      \Gamma_{11}^{2} - \hat\Gamma^{2} +
      4 \gamma_{*} C_{12} \,
      \hat\rho_{11}\hat\rho_{12}\hat\rho_{21}\hat\rho_{22}
      \cos\hat\Phi
    \right]^{2} < 0, \\
    {\mathcal{P}}_4(\Gamma_{12}) & = & - \frac{1}{4}\left[
      \Gamma_{12}^{2} - \hat\Gamma^{2} +
      4 \gamma_{*} C_{12} \,
      \hat\rho_{11}\hat\rho_{12}\hat\rho_{21}\hat\rho_{22}
      \cos\hat\Phi
    \right]^{2} < 0.
    \end{eqnarray*}
This means that there is at least one real
    root located in each of the intervals
    $P_1 \in [\Gamma_{11},\tilde{\Gamma}]$ and $P_2 \in
     [\tilde{\Gamma},\Gamma_{12}]$.

As a corollary we get that ${\mathcal{P}}_{4}(\Gamma) > 0$ for $
P_{1} < \Gamma < P_{2}$. Concerning the other two roots $Q_{1,2}$
($Q_2>Q_1$ if real), nothing may be said in general. However, for
the case ${\mathcal{P}}_* > 0 $ it is clear that
${\mathcal{P}}_4(\pm \infty)
> 0$ and then $Q_{1,2}$ are real numbers. When ${\mathcal{P}}_* < 0$ it happens
that $Q_1,Q_2$ could be  (and, in fact, they are) complex
conjugate roots for certain parameter ranges. In any case, the
boundedness of $\Gamma \in [P_1,P_2]$ ensures the periodicity of
the solutions.

\begin{figure}
\epsfig{file=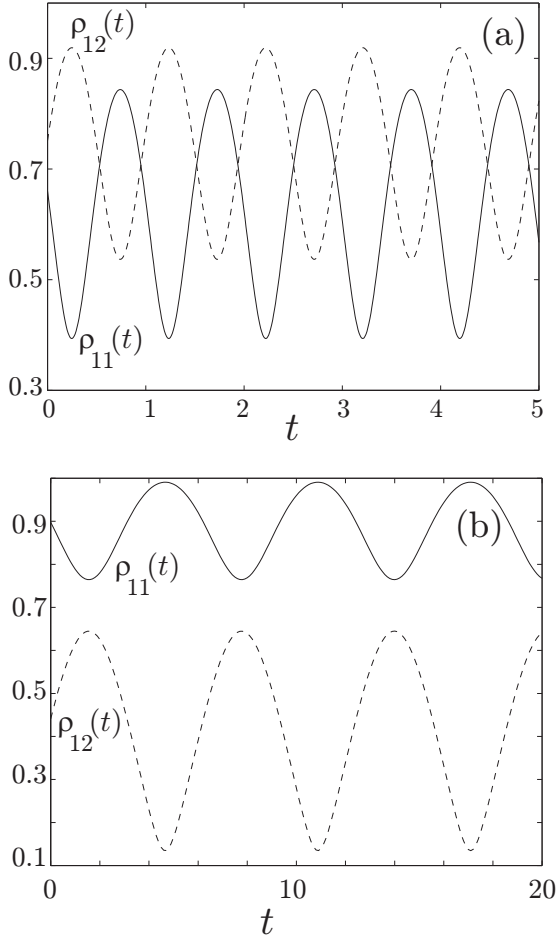,width=7.3cm} \caption{Evolution of the
densites $\rho_{11}(t)$ (solid) and $\rho_{12}(t)$ (dashed) for
two set of combinations of parameters and initial conditions (a)
$g=100$, $a_{11} = 1$, $a_{12} = 0.97$, $a_{22} = 0.94$, $\beta =
N_1/N_2 =1$, $\rho_{11}(t_0) = 0.78, \phi_{11}(t_0) = 0$,
$\phi_{12}(t_0) = 0.5, \rho_{21}(t_0) = 0.5$, $\phi_{21}(t_0) =
1.5, \phi_{22}(t_0) = 0.7$. (b) $g=100$, $a_{11} = 1, a_{12} =
0.1$, $a_{22} = 0.02$, $\beta = N_1/N_2 =1/4$, $\rho_{11}(t_0) =
0.9$, $\phi_{11}(t_0) = 0$, $\phi_{12}(t_0) = 0$, $\rho_{21}(t_0)
= 0.9$, $\phi_{21}(t_0) = 1.5$, $\phi_{22}(t_0) = 0$. We fix
${\tilde{t}} = 0$ and get the corresponding (irrelevant) value for
$t_0$ from Eq. (\ref{ttilde}).} \label{prima}
\end{figure}

Let us now proceed to find an explicit form for the solutions.
First, to present polynomial ${\mathcal{P}}_{4}(\Gamma)$ in the
canonical form we make the transform:

\begin{equation}
  \Gamma(t) = { \alpha Y + \beta \over Y + \delta},
  \label{solu}
\end{equation}
where
\begin{eqnarray}
  \alpha &=&
      \tinyhalf \left( \delta + 1 \right) P_{2}
    - \tinyhalf \left( \delta - 1 \right) P_{1},
  \\
  \beta &=&
      \tinyhalf \left( \delta + 1 \right) P_{2}
    + \tinyhalf \left( \delta - 1 \right) P_{1},
  \\
  \delta & = &
  {
  \sqrt{
    \left| P_{1} - Q_{1} \right|
    \left| P_{1} - Q_{2} \right|
  }
  +
  \sqrt{
    \left| P_{2} - Q_{1} \right|
    \left| P_{2} - Q_{2} \right|
  }
  \over
  \sqrt{
    \left| P_{1} - Q_{1} \right|
    \left| P_{1} - Q_{2} \right|
  }
  -
  \sqrt{
    \left| P_{2} - Q_{1} \right|
    \left| P_{2} - Q_{2} \right|
  }
  }.
\end{eqnarray}
Equivalently we may also write
\begin{equation}
\Gamma(t) =   P_{0} + { \Delta \over \delta } \; { Y \over \delta
+ Y }
\end{equation}
with the quantities $\Delta$ and $P_{0}$ being given by
\begin{eqnarray}
  \Delta &=& \alpha\delta - \beta =
    \tinyhalf \left( \delta^{2} - 1 \right) \left( P_{2} - P_{1}
    \right),
  \\
  P_{0} &=& { \beta \over \delta } =
  {1 \over 2\delta}
  \left[
    \left( \delta + 1 \right) P_{2} + \left( \delta - 1 \right) P_{1}
  \right],
\end{eqnarray}
It is clear that   $P_{1} < P_{0} < P_{2}$ and thus
${\mathcal{P}}(P_0)
>0$. In terms of $Y$ the polynomial ${\mathcal{P}}_{4}(\Gamma)$ can be
written as
\begin{equation}
  {\mathcal{P}} (\Gamma) =
  \delta^{4} {\mathcal{P}}(P_{0})
  \;
  {
    \left( 1 - Y^{2} \right)
    \left( 1 - k^{2} Y^{2} \right)
  \over
    \left( \delta + Y \right)^{4}
  }.
\end{equation}
Thus, the function $Y(t)$ satisfies the equation
\begin{equation}
  Y_{t} =
  \widetilde\Omega \;
  \sqrt{
    \left( 1 - Y^{2} \right)
    \left( 1 - k^{2} Y^{2} \right)
  }
\label{dYdt}
\end{equation}
where
\begin{eqnarray}
  \widetilde\Omega & = &
  {\delta^{2} \over \Delta } \sqrt{ {\mathcal{P}}(P_{0}) }
  \nonumber \\
  &=&   { 1 \over 2 }
  \sqrt{ \left| {\mathcal{P}}_{*} \right| }
  \left[
    \sqrt{
    \left| P_{1} - Q_{1} \right|
    \left| P_{2} - Q_{2} \right|
    }
  +
    \sqrt{
    \left| P_{1} - Q_{2} \right|
    \left| P_{2} - Q_{1} \right|
    }
  \right].
\end{eqnarray}
Equation (\ref{dYdt}) can be solved as follows:
\begin{equation}
\label{snsn}
  Y(t) = {\sn} \left( \widetilde\Omega \left(t - \hat t\right), k
  \right),
\end{equation}
where
\begin{equation}
  k =
  {
  \sqrt{
    \left| P_{1} - Q_{1} \right|
    \left| P_{2} - Q_{2} \right|
  }
  -
  \sqrt{
    \left| P_{1} - Q_{2} \right|
    \left| P_{2} - Q_{1} \right|
  }
  \over
  \sqrt{
    \left| P_{1} - Q_{1} \right|
    \left| P_{2} - Q_{2} \right|
  }
  +
  \sqrt{
    \left| P_{1} - Q_{2} \right|
    \left| P_{2} - Q_{1} \right|
  }
  }
\end{equation}
and the constant $\hat t$ may be determined from the initial
conditions by solving the equations
\begin{equation}
\label{ttilde}
  \hbox{sn}\left( \widetilde\Omega \left(t_{0} - {\hat t}\right), k \right) =
  { \Delta \over \alpha - {\hat\Gamma}} - \delta.
\end{equation}
This allows us to get a explicit solution for our problem. To do
so let us first use  Eq. (\ref{snsn}) to find
\begin{equation}
\Gamma(t) = \frac{\alpha  {\sn} \left( \widetilde\Omega \left(t -
{\hat t}\right), k \right) + \beta}{ {\sn} \left( \widetilde\Omega
\left(t - {\hat t}\right), k \right)+ \delta}. \label{finalsolu}
\end{equation}
 Thus, inserting this explicit expression for $\Gamma(t)$ and Eqs.
 (\ref{rsol11})-(\ref{rsol22}) one may get explicit expressions
 for the populations $\rho_{jk}(t), j,k=1,2.$ Typical forms of the
 solutions for different parameter values are shown in Fig.
 \ref{prima}.

 The period of oscillations of the populations $\rho_{jk}(t)$ is
 that of the sn function, i.e.
\begin{eqnarray}
  T & =& { 4K(k) \over \widetilde\Omega } =
  {
    4\Delta K(k)
  \over
    \delta^{2} \sqrt{{\mathcal{P}}(P_{0})}
  } = \frac{8 K(k)}{\sqrt{ \left|  {\mathcal{P}_{*}} \right| }} \times \nonumber \\
 &  & {1
  \over
    \sqrt{
      \left| P_{1} - Q_{1} \right|
      \left| P_{2} - Q_{2} \right|
    }
    +
    \sqrt{
      \left| P_{1} - Q_{2} \right|
      \left| P_{2} - Q_{1} \right|
    }
  }
\end{eqnarray}
where the quantity $K(k) = \int_0^{\pi/2}
d\theta/\sqrt{1-k^2\sin^2 \theta}$ is the complete elliptic
integral of first kind \cite{Abramowitz}.

Concerning the period, the dependence on the physical parameter
$N$ is easy to obtain. Let us define the new constants (not
dependent on $N$) $\bar{\gamma}_{ij} = \gamma_{ij}/N$,
${\bar{u}}_{ij} = u_{ij}/N$, ${\bar{\Gamma}}_{ij} =
\Gamma_{ij}/N$, ${\bar{\gamma}}_* = \gamma_*/N^2$. Using these
definitions we find that the roots $P_1,P_2,Q_1,Q_2$ scale
linearly on $N$. Thus $k,\delta$ and ${\cal{P}}_*$ do not depend
on $N$ and the period scales as $T(k,N) = N^{-1} T(k,1)$.

\begin{figure}
\epsfig{file=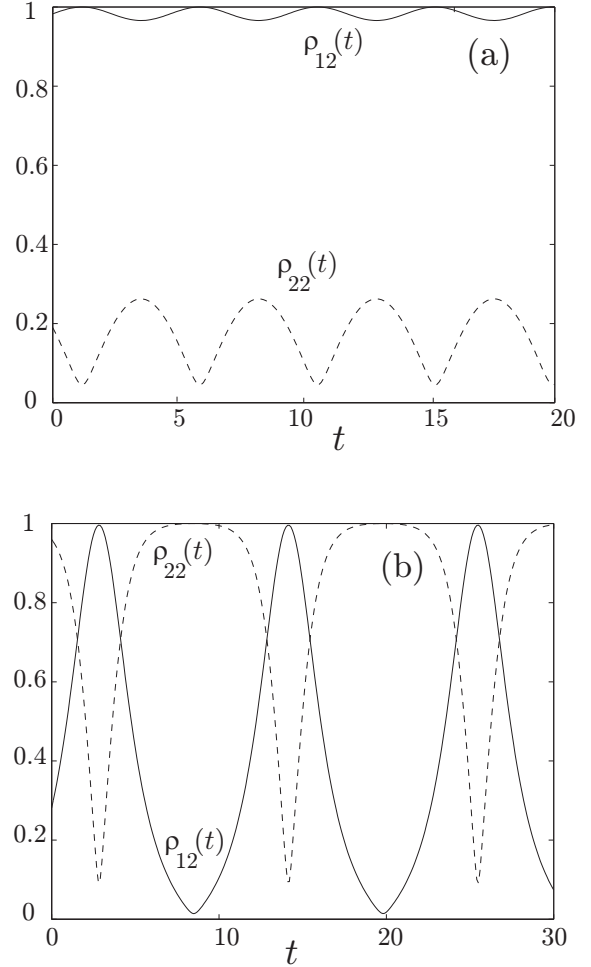,width=7.5cm} \caption{Evolution of the
densites $\rho_{12}(t)$ and $\rho_{22}(t)$ (containing the
vorticity of each component) for two set of combinations of
 initial conditions corresponding respectively to (a) a stable
 vortex in $|1>$ and (b) an unstable vortex in
 $|2>$. Parameter values are $g=100$,  $\beta = N_1/N_2 =1$, $a_{11} =
 1$,
 $a_{12} = 0.97$, $a_{22}= 0.94$.  Initial conditions are (a)
$\rho_{11}(t_0) = 0.001$, $\phi_{11}(t_0) = 0$, $\phi_{12}(t_0) =
0.1$, $\rho_{21}(t_0) = 0.999$, $\phi_{21}(t_0) = 0$,
$\phi_{22}(0) = 0.2$.(b) $\rho_{11}(t_0) = 0.9999$,
$\phi_{11}(t_0) = 0$, $\phi_{12}(t_0) = 0.1$, $\rho_{21}(t_0) =
0.001$, $\phi_{21}(t_0) = 0$, $\phi_{22}(t_0) = 0.2$. We fix
${\tilde{t}} = 0$ and get the corresponding (irrelevant) value for
$t_0$ from Eq. (\ref{ttilde}).} \label{dual}
\end{figure}

Concerning the instability mechanisms described in Refs.
\cite{PRL3,PRA3}, they are now completely understood within the
framework of Eq. (\ref{finalsolu}). For instance in Fig.
\ref{dual} we plot typical solutions in the regime of vortex
trasfer for parameter values typical of the experiments
\cite{Scattering-lengths}. Both the stability of the vortex when
it is placed in $\left| 1\right>$ and its instability when put on
$\left| 2\right>$ are well described by the exact solutions of the
two--mode model.

\section{Conclusions}
\label{conclu}

 In this paper we have integrated the two--mode model developed
 for the explanation of the vortex transfer mechanisms described in
 Refs. \cite{MMatt99,PRL3,PRA3}. The most remarkable result is
 that all the solutions are periodic functions and thus the
 transfer mechanisms of at least part of the vorticity are natural
 within the range of validity of
 the model, which was previously found to be at least that of
 pancake traps and some regimes of fully three-dimensional traps.
 It has been shown that the frequency of the oscillations depends
 linearly on the total number of particles of the condensate while
 the dependence on the other parameters (relative populations and
 scattering lengths)
 is nontrivial and given by our explicit formulae.

 We hope that the technique described here will be useful to
 analyze the similar problem which arises when a Josephson
 coupling  between both species is incorporated into
 the experimental setup by means of a off-resonant laser
 \cite{Sols}.

\acknowledgements

We are grateful to J. J. Garc\'{\i}a-Ripoll for discussions. This
work has been partially supported by the Ministerio de Ciencia y
Tecnolog\'{\i}a under grant BFM2000-0521. V. Vekslerchik is
supported by Ministerio de Educaci\'on under grant SB99-AH777133.


\end{document}